\documentclass[preprint,12pt,showpacs]{revtex4} 
\usepackage{amsmath}
\usepackage{latexsym}
\usepackage{amssymb}
\usepackage{graphics,epstopdf}
\usepackage{amsmath,amssymb,amsthm}
\usepackage{slashed}
\usepackage[colorlinks=true, citecolor=blue, urlcolor=blue ]{hyperref}

\begin{document}
\title{On the position dependent effective mass Hamiltonian }
\author{Kalpana Biswas}
\email{klpnbsws@gmail.com}
\affiliation{Department of Physics, University of Kalyani, West Bengal, India-741235}
\affiliation{Department of Physics, Sree Chaitanya College, Habra, North 24 Parganas, West Bengal-743268}
\author{Jyoti Prasad Saha}
\email{jyotiprasadsaha@gmail.com}
\affiliation{Department of Physics, University of Kalyani, West Bengal, India-741235}
\author{Pinaki Patra}
\thanks{Corresponding author}
\email{monk.ju@gmail.com}
\affiliation{Department of Physics, Brahmananda Keshab Chandra College, Kolkata, India-700108}

\date{\today}

\begin{abstract}
Noncommutivity of position and momentum makes it difficult to formulate the unambiguous structure of the kinetic part of Hamiltonian for the position-dependent effective mass (PDEM).  Various existing proposals of writing the viable kinetic part of the Hamiltonian for PDEM, conceptually lack from first principle calculation.  Starting from the first principle calculation, in this article, we have advocated the proper self-adjoint form of the kinetic part of Hamiltonian for PDEM. We have proposed that ambiguity of construction of viable kinetic part for PDEM can be avoided if one takes the care from the Classical level combination of position and momentum.  \\
In the quantum level, the spatial points do not appear in equivalent footing for the measure of inertia (mass). This exhibits the existence of an inertia potential. Thus the new structure of the Kinetic part differs from the existing structure of the kinetic part of Hamiltonian by providing an extra potential like contribution.  This inertia potential can be absorbed with the external potential and redefine the known structure of PDEM under this effective potential. This enables us to apply the existing formalism of quantum mechanics. The coherent state structures for the newly proposed form of Hamiltonian are provided for a few simple experimentally important models.
\end{abstract}

 \maketitle
\section{Introduction}
Position dependent effective mass (PDEM) was first considered in the description of electronic properties and band structure of semiconductor Physics \cite{CostaFilho,Muharimousavi,Souza Dutra,Souza Dutra1,Schmidt,Abdalla,Jha,Jha1,heterostructure1,heterostructure2,Mario}. Later it was proved to be a useful concept in other branches of Physics. Efficacious application of the concept of PDEM in various branches of Physics had made the concept of PDEM a tropical issue \cite{yu,lozada,arias,guedes,mello,santos,cavalcanti,cunha,bekke,vitoria,vitoria1,bekke1}. However, due to the noncommutativity of PDEM and momentum, it was realized from the early days of the PDEM, that it is not straightforward to write the self-adjoint Hamiltonian for this type of system. It was realized that the kinetic part of the Hamiltonian will be of the following form (for simplicity, we are considering one-dimensional case)\cite{heterostructure1,heterostructure2}.
\begin{equation}\label{heterostructurehamiltonian}
\hat{H}= -\frac{1}{2} m^\alpha \left(\frac{d}{dx}\right)m^\beta \left(\frac{d}{dx}\right)m^\alpha .
\end{equation}
With the constraint on the constant parameters $\alpha$ and $\beta$
\begin{equation}\label{constraintalpha}
\beta+2\alpha =-1.
\end{equation}
To fix the values of $\alpha$ and $\beta$, one had to rely on some known solvable models. For example, photoluminescence excitation spectra from $GaAs/Al_{0.35}Ga_{0.65}As$ had suggested that $\beta \approx -1$ is a potential candidate. Later it was suggested by various theoretical and experimental models that $\alpha=0,\; \beta=-1$ and this condition is called Bastard's condition. However, experimentally, it was further shown that the results of $GaAs/Al_{x}Ga_{1-x}As$ and $Si-Ge$ heterojunction is insensitive to the values of $\beta$ and even $\beta=0$ satisfies the results. Therefore, the kinetic part of the Hamiltonian of PDEM, demands careful construction. All the available literature in this regime deals with the problem of noncommutativity between position-dependent mass and momentum by Weyl quantization restricting the product of mass and momentum only with the combination of powers of mass followed by the momentum which is further followed by powers of mass and momentum and finally a mass term. However, the further combination appears if one considers the combination of products from the Classical level of Legendre transformation between Lagrangian and Hamiltonian. In particular, $\frac{1}{m}pp+pp\frac{1}{m}$ is an allowed combination which is ignored so far. In this article, we have precisely substantiated this job. Considering the problem from the first principle calculation we have overtured a viable kinetic part of the Hamiltonian for the position-dependent mass problem. \\
The constitution of the article is the following, In the next section, the kinetic part for PDEM Hamiltonian is constructed from the first principle calculation. Construction of Coherent state structure with the help of the formalism of supersymmetric Quantum mechanics is an effective method to solve the PDEM problem. This is described in section-III. At last, the conclusion and future aspects are described. 
\section{Construction of the proper structure of PDEM Hamiltonian}
For a system with $n$ degrees of freedom, if $X=(x_1, x_2,...., x_n)^T$ is the generalized co-ordinate vector, then the all Classical dynamical information about the system are encoded in the Lagrangian functional \cite{gantmacher,arnold}
\begin{equation}\label{lagrangian1}
\mathcal{L}\left(x_1,x_2,..,x_n,\dot{x}_1,\dot{x}_2,..\dot{x}_n,t \right) =\frac{1}{2}\dot{X}^T M \dot{X} - V\left( x_1,x_2,...x_n,t\right) .
\end{equation}
Here, dot$(.)$ denotes the derivative with respect to the parameter time $t$ and $X^T$ stands for transpose of $X$ in the sense of ordinary matrix operation. The potential function $V$ in general may depend on the velocities $(\dot{x}_i, \; i=1,..n)$, but we shall consider the cases where it can only be the function of co-ordinates. The positive definite matrix $M$ is called inertia matrix or simply the mass matrix.\\
Equivalently, one can describe the systems by the energy functional, called Hamiltonian, which is the Legendre transformation of the Lagrangian as follows.
\begin{equation}\label{legendre1}
\mathcal{H}\left( x_1,x_2,...x_n,,p_1,p_2,...,p_n,t\right)=\sup_{\{\dot{x}_1, \dot{x}_2,..\dot{x}_n\}}\left(\Sigma_{i} (p_i \dot{x}_i)-\mathcal{L} \right).
\end{equation}
From the extremization condition in \eqref{legendre1} over $\dot{x}_i$, one gets the conjugate momentum $p_i$ corresponding to the generalized co-ordinate $x_i$ as $p_i=\frac{\partial \mathcal{L}}{\partial \dot{x}_i}$. 
All the classical observables can be expressed as the functions of the positions, momenta and time. For example, in the one-dimensional case, the energy can be expressed by the Hamiltonian function
\begin{equation}
H(x,p)=\frac{1}{2m}p^2 + V(x).
\end{equation}

In Quantum theory, the observables are self-adjoint operators acting on some Hilbert space (depending upon the system under consideration). Better or worse, the strategy of composing the Quantum theory corresponding to a known Classical system (i.e, the system for which the classical observables are known), one constructs Quantum observables corresponding to the Classical observables by Bohr's correspondence principle. Being the operators acting on the Hilbert space corresponding to the system, the quantum observables may not commute to each other. The knowledge on the commutation relations among the concerned observables is necessary for the description of a Quantum theory.  For example, the commutation relations for the position operators $(x_i)$ and the momentum operators $(p_i)$ are given by
\begin{equation}
[x_i,p_j]=i\hbar \delta_{i,j},\;\; [x_i,x_j]=[p_i,p_j]=0.
\end{equation}
Here, $\delta_{i,j}$ is Kroneker delta whose values are $1$ for equal values of indices and $0$ otherwise. $\hbar$ is Plank constant which determines the natural restriction on our inability for the concurrent sharp measurement of position and momentum operators. \\
It is evident that for the effective mass theory, the effective mass can be a function of position and time. Therefore, care should be taken from the very first line of construction of a viable effective mass theory. Although the effective mass occurs only in the domain of quantum theory, let us assume that there exists a classical model, quantization of which is the resultant quantum theory of effective mass. 
Whether we write $x_i$ after $p_i$ or $p_i$ after $x_i$, is immaterial in Classical case. In quantum theory, it is a serious issue.  In this respect, for the case of PDEM, we are advocating for the amendment of redefining \eqref{legendre1} by replacing $p_i\dot{x}_i$ with   $\frac{1}{2} (\dot{x}_ip_i + p_i\dot{x}_i)$ and replace $\dot{x}_i$ by $\frac{1}{m}p_i$ and $p_i\frac{1}{2}m$ in the classical level calculation.  The ordering of observables (in classical level) might be set forth to equal footage in this fashion. No special ordering will be preferred. This will commit the improve combination of observables in Hamiltonian which can directly be promoted to Quantum theory. The construction is as follows.\\
For convenience, let us consider one generalized coordinate $x$ and its conjugate momentum $p$. For the Lagrangian $L(x,\dot{x},t)=\frac{1}{2}\dot{x}m(x,t)\dot{x}-V(x,t)$, the classical Hamiltonian is considered to be given by`
\begin{equation}\label{lagrangiantohamiltonian}
H(x,p,t)=\sup_{\{\dot{x}\}}\left[\frac{1}{2}(\dot{x}p+p\dot{x})-L\right] = \sup_{\{\dot{x}\}}[\mathcal{L}_{\mathcal{H}}(x,\dot{x},p,t)].
\end{equation}
Supremum attains for the condition
\begin{equation}\label{legendreextremum}
\frac{\delta\mathcal{L}_{\mathcal{H}}}{\delta \dot{x}}=0= \left(\frac{1}{2}p + \frac{1}{2}p \right)-\left(\frac{1}{2}m\dot{x} + \frac{1}{2}\dot{x}m \right).
\end{equation}
Now, we shall take $p=m\dot{x}\Rightarrow \dot{x}=\frac{1}{m}p$ in the first term of the Legendre transformation \eqref{lagrangiantohamiltonian} as well as for the leftmost $\dot{x}$ of Lagrangian and for other $\dot{x}$ in \eqref{lagrangiantohamiltonian} we shall use $p=\dot{x}m\Rightarrow \dot{x}=p\frac{1}{m}$. This leads to \eqref{classical hamiltonian}. 
In particular, in one dimension case, the Hamiltonian reads (See \eqref{lagrangiantohamiltonian} and \eqref{legendreextremum}) as follows
\begin{equation}\label{classical hamiltonian}
\mathcal{H}=\frac{1}{2}\left( \frac{1}{m}pp+pp\frac{1}{m}\right)- \frac{1}{2m}pmp\frac{1}{m} +V(x,t).
\end{equation}
Upon quantization, one can have the quantum observables $\hat{p}$ and $\hat{x}$, corresponding to the classical variable $p$ and $x$ respectively. In $\{\vert x\rangle\}$ representation $\hat{p}=\frac{1}{i}\frac{\partial}{\partial x}$ (Using $\hbar=1$) and the correct Hamiltonian for arbitrary position and time dependent mass $(m(x,t))$ reads
\begin{equation}\label{quantum hamiltonian}
\hat{H}=-\frac{\partial}{\partial x}\left(\frac{1}{2m}\frac{\partial}{\partial x}\right) - \frac{1}{2m}\left(\frac{m'}{m}\right)^2 + V(x,t).
\end{equation} 

Here, prime denotes the derivative with respect to $x$.
The Hamiltonian in \eqref{quantum hamiltonian} is Hermitian only for real-valued function $m(x,t)$. However, the dynamic effective mass of granular media may be complex-valued (frequency-dependent).  If, $m(x,t)$ is complex, then the correct expression may be written by
\begin{equation}\label{complexmasshamiltonian}
\hat{H}= -\frac{1}{4}\frac{\partial}{\partial x}\left(\left(\frac{1}{m} + \frac{1}{m^*}\right)\frac{\partial}{\partial x}\right) - \frac{1}{4}\left(\frac{(m_x)^2}{m^3} + \frac{(m^*_x)^2}{(m^*)^3}\right) + V(x,t).
\end{equation}
Here, suffix $x$ stands for derivative with respect to $x$.\\
The form of Hamiltonian \eqref{quantum hamiltonian} agrees with \eqref{heterostructurehamiltonian} for $\alpha=0, \beta=-1$ case, except for the extra position and time-dependent function $ V_{inertia} = \frac{1}{2m}  \left(\frac{m'}{m}\right)^2 $ which can be considered as potential function arise due to the PDEM. The occurrence of this intrinsic potential function is natural. Because, if we recall that, the concept of mass is nothing but a measure of inertia. Now, if the measure of inertia becomes a function of the spatial variables, then automatically not every point in space is in equal footing in the sense of inertia of the object. That's why there will be some balancing natural potential. The same argument is true for the complex mass case. It is worth noting that, for the case of abrupt heterostructure \eqref{heterostructurehamiltonian}, the natural inertia potential arises as
\begin{equation}
V_{inertia}= -\frac{\alpha}{2}\left\{ (\beta+\alpha - 1)\frac{m'^2}{m^3} + \frac{m''}{m^2} \right\}.
\end{equation}
This, of course, vanishes for $\alpha=0$. \\
Recently, complex-valued PDEM had drawn attention due to its appearance in jammed granular materials in which PDEM maybe even frequency-dependent. The Hamiltonian, in that case, differs from the form of Hamiltonian of real-valued PDEM, only in the expression of the effective potential. In particular,
\begin{equation}\label{complexpdemhamiltonian}
\mathcal{H}^c=-\frac{1}{2}\frac{d}{dx}\left(\frac{1}{\mu}\frac{d}{dx}\right) + V^c_{\mbox{eff}}.
\end{equation}
With
\begin{eqnarray}
V^c_{\mbox{eff}}=V-\frac{1}{4}\left(\frac{m'^2}{m^3}+ \frac{(m^*)'^2}{(m^*)^3}\right).\\
\frac{1}{\mu}=\frac{1}{m} + \frac{1}{m^*}.
\end{eqnarray}

In the next section, the construction of the coherent state structure for real PDEM is described. 
\section{Construction of Coherent State for real PDEM}
To construct the coherent states, one can use the algorithm of Supersymmetric (SUSY) quantum mechanics \cite{coherent1,coherent2,coherent3,coherent4,coherent5,coherent6,coherent7,coherent8,coherent9,coherent10,coherent11,coherent12}, in which one seeks an intertwining relation between the original potential and its supersymmetric partner potential by the relation
\begin{equation}\label{intertwining}
\hat{A}\hat{H}=\hat{\tilde{H}}\hat{A}.
\end{equation}
Where 
\begin{eqnarray}
\hat{H}=-\frac{1}{2m(x)}\frac{\partial^2}{\partial x^2} + \frac{m'}{2m^2}\frac{\partial}{\partial x} +V_{\mbox{eff}},\\
\mbox{With}\;\;\; V_{\mbox{eff}} = V-\frac{m'^2}{2m^3}.\\
\hat{\tilde{H}}=-\frac{1}{2m(x)}\frac{\partial^2}{\partial x^2} + \frac{m'}{2m^2}\frac{\partial}{\partial x} +\tilde{V}.
\end{eqnarray}
Using \eqref{intertwining}, one can construct the annihilation operator by assuming the form
\begin{equation}\label{annihilation}
\hat{A}= \frac{1}{\sqrt{2}}\left( a(x)\frac{d}{dx}a(x) +\phi(x)\right).
\end{equation} 
To verify, whether $\hat{A}$ is an annihilation operator, one can consider its commutation relation with its adjoint operator (creation operator) as follows.
\begin{equation}
\left[\hat{A},\hat{A}^\dagger \right] =\frac{\phi'}{\sqrt{m}}.
\end{equation}
Which confirms that $\hat{A}$ is an anihilation operator for the deformed space consists of deformed position $(\hat{\phi})$ and deformed  momentum $(\hat{\Pi})$ operators which are connected with $\hat{A}$ and $\hat{A}^\dagger$ as follows.
\begin{eqnarray}
\hat{\phi}=\frac{1}{\sqrt{2}}\left(\hat{A}+\hat{A}^\dagger\right),\\
\hat{\Pi}=\frac{-i}{\sqrt{2}}\left(\hat{A}-\hat{A}^\dagger\right).\\
\therefore \left[\hat{\phi},\hat{\Pi}\right]= i\frac{\phi'}{\sqrt{m}}.
\end{eqnarray}   
Now, using \eqref{quantum hamiltonian}, \eqref{intertwining} and \eqref{annihilation}, if we apply \eqref{intertwining} on some function $\psi$ and equate the coefficients of $\frac{d^k\psi}{dx^k},\; k=0,1,2,3$, we get the following Ricatti equation from which one can get the desired $\phi(x)$.
\begin{equation}\label{RicattiK}
K'+ m\left(\frac{1}{m}\right)'K- K^2+ 2m\left( V-\frac{m'^2}{2m^3} \right) =2 \lambda m .
\end{equation}
With 
\begin{eqnarray}\label{phifromk}
\phi(x)=Ka^2-aa', \\
 a(x)=\frac{1}{m^{\frac{1}{4}}},\\
\lambda=\mbox{constant}, \nonumber\\
\tilde{V}=V_{\mbox{eff}}+\frac{1}{\sqrt{m}}\phi' .
\end{eqnarray}
For a quick consistency check, one can see that for the constant mass case, the harmonic oscillator potential $(\frac{1}{2}m\omega^2 x^2)$, leads to $ \phi = \sqrt{m}\omega x$ and $\lambda=\frac{1}{2}\omega$ which are desired. One must admit that the reduction of the original problem (a second-order linear ordinary differential equation) to Ricatti equation (1st order nonlinear ordinary differential equation) is just treating the problem in another fashion. It doesn't reduce the level of difficulty in solving the original problem. However, the intertwining factorization technique helps to get a glimpse of coherent states which is illustrated below.\\
Since
\begin{equation}
\hat{A}^\dagger \hat{A}= \hat{H} -\lambda,
\end{equation}
one can conclude that the states which are eigenfunctions of our constructed annihilation operator $\hat{A}$, are the coherent states of the PDEM Hamiltonian \eqref{quantum hamiltonian}. That means, if 
\begin{equation}
\hat{A}\vert\alpha\rangle =\alpha \vert\alpha\rangle\;,\;\; \alpha \in \mathbb{C},
\end{equation}
then $\vert\alpha\rangle$ is the coherent state of $\hat{H}$. One can indeed verify that $\vert\alpha\rangle$ satisfies the minimum uncertaintity condition (the necessary condition to be coherent state). 
It can be shown in straightforward manner that
\begin{equation}\label{dispersionphipi}
\left(\Delta\phi\right)_{\vert\alpha\rangle}^2 = \left(\Delta\Pi\right)_{\vert\alpha\rangle}^2 = \frac{1}{2} \langle \alpha\vert \frac{\phi'}{\sqrt{m}}\vert\alpha \rangle .
\end{equation}
That means, the state $\vert\alpha\rangle$ is such state which minimizes the uncertainty condition. In particular,
\begin{equation}
\left(\Delta\Phi\right)_{\vert\alpha\rangle}^2 \left(\Delta\Pi\right)_{\vert\alpha\rangle}^2 \ge \frac{1}{4} \vert\langle\alpha \vert \left[\Phi, \Pi\right]\vert \alpha\rangle \vert^2 .
\end{equation}
To write down the exact amount of uncertainty, one needs the explicit form of $\vert\alpha\rangle$ which can be constructed by constructing the eigenfunction of the annihilation operator $\hat{A}$.\\

For a demonstration, let us take the PDEM of the form
\begin{equation}\label{pdemformreal}
m(x)=m_0e^{\nu x}.
\end{equation}
Constant mass $(m_0)$ case can be revealed by setting the parameter $\nu \rightarrow 0$. This form of mass is not just for mathematical simplicity, rather it corresponds to the description of transport regime in a semiconductor superlattice characterized by extreme anisotropy of effective mass \cite{Mario}. The experimental results corresponding to $HgTe$ and  $Hg_{0.65}Cd_{0.35}Te$ were slightly differed from the linear mass approximation (See Fig.2 of \cite{Mario}). Therefore, the effective mass function of \eqref{pdemformreal} may be suitable candidate for the effective mass tailoring of electron in $HgTe$ and  $Hg_{0.65}Cd_{0.35}Te$.\\
Let us choose the potential function to be a simple potential well of the form
\begin{eqnarray}\label{potentialwell}
V(x)=\left\{ \begin{array}{cc}
-V_0^2, & 0\le x\le l\\
0 & \mbox{Otherwise}.
\end{array}\right.
\end{eqnarray} 
Because of the high sensitivity with the parameter values of the nonlinear equations (such as \eqref{RicattiK}), it is not difficult to understand that it will be easier to get the closed-form solution for specific choices of parameters. We set the parameter
\begin{equation}\label{parametervalues}
 \nu=-12.
\end{equation}
The choice of the parameter value $\nu=-12$ is purely for mathematical convenience. The choice $\nu=-12$ helps to express the solution of the concerned Ricatti equation (RE) in terms of the Bessel function of the first kind ($J_{\frac{1}{2}}$ and $J_{-\frac{1}{2}}$).\\
The conceptual part of the construction of coherent states is easy. Whereas, being a nonlinear differential equation, the construction of the solutions of the RE is not straightforward. Indeed, no method can be utilized to construct a complete set of solutions for RE. 
However, there are some schemes, by which one can construct a class of solutions for RE. Such an elegant method is given in \cite{ricatti}. First, let us recall the general form of RE \cite{ricatti,ricatti1,ricatti2,ricatti3,ricatti4,ricatti5,ricatti6,ricatti7,ricatti8}, which is given by
\begin{equation}\label{generalricatti}
\frac{dK}{dx}+P(x)K+Q(x)K^2=R(x).
\end{equation}
Where $P(x)$, $Q(x)$ and $R(x)$ are arbitrary functions of $x$. The  nonlinearity is quadratic in $K(x)$.
Let us introduce the following transformation of $K(x)$.
\begin{equation}\label{canonicallike}
\bar{K}(x)=f(x)e^{\int^x g(x_1)K(x_1)dx_1}.
\end{equation}
The functions $f(x)$ and $g(x)$ have to be determined from the consistency conditions, which give the following set of relations.
\begin{eqnarray}
2\frac{f'}{f}+\frac{g'}{g}=P, \\
g=Q ,\\
\bar{K}''=gS\bar{K}. \label{schrodingerricatti}\\
\mbox{With}, \; S(x)=R+\frac{f''}{fg}.
\end{eqnarray}
Therefore,  with the help of the canonical transformation ~\eqref{canonicallike}, one can reduce the RE into a Schr\"{o}dinger equation ~\eqref{schrodingerricatti}.
If one can solve ~\eqref{schrodingerricatti}, then the solution $\bar{K}$ can be utilized to construct the solution of RE by the following differential form of the canonical transformation ~\eqref{canonicallike}.
\begin{equation}\label{k from k'}
K(x)=\frac{1}{g}\frac{d}{dx}\ln\left(\frac{\bar{K}}{f}\right).
\end{equation}
Now for the specific mass \eqref{pdemformreal} and the potential \eqref{potentialwell}, in the region beyond $0\le x\le l$, the RE takes form
\begin{equation}\label{explicitricatti}
K'-\nu K-K^2=\nu^2 + 2\lambda m_0e^{\nu x}.
\end{equation}
Comparing ~\eqref{explicitricatti} with ~\eqref{generalricatti}, one can identify that
\begin{eqnarray}
P=-\nu,\;\; Q=-1,\;\; R=\nu^2+2\lambda m_0e^{\nu x}.\\
\therefore g(x)=-1, \;\; f(x)=f_0e^{-\frac{\nu}{2}x}.\\
\mbox{And}, \;S(x)=\frac{3}{4}\nu^2 -2V_0^2 m_0 e^{\nu x}.
\end{eqnarray}
For convenience, let us change the variable
\begin{equation}
z=-\frac{3}{4}\nu+\frac{3}{4}\nu^2 + 2\lambda m_0e^{\nu x}.
\end{equation}
Then ~\eqref{schrodingerricatti} takes the form
\begin{equation}\label{schrodingerchange}
\frac{d^2 \bar{K}}{dz^2}+\frac{1}{z}\frac{d \bar{K}}{dz}+\left(\frac{1}{\nu^2 z}+ \frac{3}{4\nu z^2}\right)\bar{K}=0.
\end{equation}
With the help of Frobenius method, one can solve eq.~\eqref{schrodingerchange}. The solution reads
\begin{equation}\label{solutionbessel}
\bar{K}(z)=c_1 \Gamma(1-\frac{\sqrt{3}}{\sqrt{\nu}}i)J_{-\frac{\sqrt{3}}{\sqrt{\nu}}i}\left( \frac{2}{\nu}\sqrt{z}\right) + c_2 \Gamma(1+\frac{\sqrt{3}}{\sqrt{\nu}}i)J_{\frac{\sqrt{3}}{\sqrt{\nu}}i}\left( \frac{2}{\nu}\sqrt{z}\right).
\end{equation}
Where $J_n(x)$ stands for the Bessel function of first kind. $c_1,c_2$ are integration constants.
For $\nu=-12$, the solution ~\eqref{solutionbessel} reduces to the linear combination of $J_{\frac{1}{2}}$ and $J_{-\frac{1}{2}}$. 
 Using ~\eqref{k from k'}, one can write down the form of $K(x)$, which reads
\begin{eqnarray}\label{Kforrealmass}
K(x)=\left\{ \begin{array}{cc} 6\left(1 - \sqrt{3}\cot\left[6\sqrt{3}\left(x+\zeta \right)\right] \right),& 0\le x\le l \\
-\frac{d}{dx}\ln \left[c_0\frac{\sin\left(\frac{1}{6}\sqrt{117-2V_0^2m_0e^{-12x}}+\theta_0 \right)}{(117-2V_0^2m_0e^{-12x})^{\frac{1}{4}}}e^{-6x} \right]  & \mbox{Otherwise}.
\end{array}\right.
\end{eqnarray}
Where $\theta_0=-\tan^{-1}\frac{2c_1}{c_2}$. 
The integration constant $\zeta$ and $\theta_0$ may be set to any value including 0 and $c_0$ can take any value except $0$ ($c_0=0$ corresponds to the trivial solution namely $\vert\alpha\rangle=0$). \\
We can note that for constant mass case $K(x)$ reduces to
\begin{equation}
\lim_{\nu\rightarrow 0}K(x)=\frac{1}{x+\zeta}.
\end{equation}
Using \eqref{Kforrealmass} in \eqref{phifromk}, the superpotential $\phi(x)$ reads
\begin{equation}
\phi(x)= -\frac{\nu}{4\sqrt{m_0}}\left( 1+ 2\sqrt{3} \cot\left[\frac{\sqrt{3}}{2}\nu\left(x+\zeta \right)\right] \right) e^{-\frac{1}{2}\nu x}.
\end{equation}
The desired coherent state then turns out to be
\begin{eqnarray}\label{coherent state of pdem1}
\vert \alpha \rangle =\left\{ \begin{array}{cc} 
\alpha_0 e^{\left(-6x- \frac{\alpha}{6}\sqrt{2m_0}e^{-6x} \right)} \sin\left[6\sqrt{3}(x+\zeta)\right], & 0\le x\le l.\\
 c_0 \frac{\sin\left(\frac{1}{6} \sqrt{117-2V_0^2m_0e^{-12x}}+\theta_0 \right)}{(117-2V_0^2m_0 e^{-12x})^{\frac{1}{4}}} e^{-6x-\frac{\alpha\sqrt{m_0}}{6}e^{-6x}},   & \mbox{Otherwise}.
\end{array}\right.
\end{eqnarray}

Better refinement of this model may be done by incorporating the following PDEM.
\begin{equation}\label{modpdem}
m(x)= m_0 e^{ax+\frac{1}{2}bx^2},
\end{equation}
For PDEM ~\eqref{modpdem}, the coherent state can be constructed for the potential ~\eqref{potentialwell}. In particular, this reads
\begin{equation}
\vert\alpha\rangle = \alpha_0 Exp\left[\kappa_1 x+ \kappa_2 x + \frac{\alpha\sqrt{m_0\pi}}{2\sqrt{2b}}e^{-\frac{a^2}{2b}}\mbox{erfi}\left[\frac{a+bx}{\sqrt{2b}}\right]\right].
\end{equation}
With
\begin{eqnarray}
\kappa_1=\frac{3}{4}a+\frac{3}{2}bc_1(a+\frac{1}{3}), \nonumber\\
\kappa_2 = \frac{3b-1}{8}+\frac{3}{2}b^2c_1, \nonumber \\
c_1= -\frac{1}{9b^2}\sqrt{\frac{1+27b^2}{3}}.
\end{eqnarray}
Normalization of $\vert\alpha\rangle$ is ensured if $\Re(\alpha)<0$ and b satisfies the relation
\begin{equation}
2187b^6 - 1458b^5 + 243 b^4-1728b^2\ge 64.
\end{equation}

\section{Conclusions}
Classical level tracking of the noncommutativity between mass (position dependent) and momentum,  provides an extra potential alike term in PDEM hamiltonian. Since, for position-dependent mass, the measure of inertia becomes position-dependent, the spatial points are not in equal footing in the sense of inertia measure. This leads to an appearance of extra potential like term which we have called "inertia potential". The extra inertial potential can be easily absorbed with the external potential term and then the new formalism proposed in this article reduced to the older form of previously known results. The construction of coherent state structure from supersymmetric quantum mechanics formalism is done for a possible model reported recently.\\
The formalism proposed in this article can be utilized for the description of various aspects of PDEM. This method can be utilized for complex PDEM without any difficulty. Several fundamental aspects, such as the entropic uncertainty principle for PDEM, the thermodynamic properties for PDEM, could be interesting for future study.

\end{document}